\documentclass[aps,prl,10pt,superscriptaddress,twocolumn,longbibliography]{revtex4-1}

\usepackage{amsmath}
\usepackage{amssymb}
\usepackage{amsfonts}
\usepackage{ascmac}
\usepackage{color}
\usepackage{comment}
\usepackage{here}
\usepackage{physics}
\usepackage{qcircuit}
\usepackage{siunitx}
\usepackage{tikz}
\usepackage{ulem}
\usepackage{tabularx}

\usepackage{hyperref}
\hypersetup{
    colorlinks=true,
    linktocpage=true,
    linkcolor=blue,
    filecolor=magenta,
    urlcolor=cyan,
    bookmarks=true,}

\usepackage{textcomp}

\definecolor{darkred}{rgb}{.7,0,0}
\newcommand{\sn}[1]{\textcolor{black}{#1}}
\newcommand{\Erase}[1]{\if0{#1}\fi}

\begin{document}


\title{\sn{Probing instantaneous quantum circuit refrigeration in the quantum regime}}

\author{Shuji Nakamura}\email{shuji.nakamura@aist.go.jp}
\affiliation{National Institute of Advanced Industrial Science and Technology (AIST), National Metrology Institute of Japan (NMIJ), 1-1-1 Umezono, Tsukuba, Ibaraki 305-8563, Japan}

\author{Teruaki Yoshioka}
\affiliation{Department of Physics, Tokyo University of Science, 1--3 Kagurazaka, Shinjuku, Tokyo 162--0825, Japan}
\affiliation{RIKEN Center for Quantum Computing (RQC), 2--1 Hirosawa, Wako, Saitama 351--0198, Japan}
\affiliation{National Institute of Advanced Industrial Science and Technology (AIST), National Metrology Institute of Japan (NMIJ), 1-1-1 Umezono, Tsukuba, Ibaraki 305-8563, Japan}

\author{Sergei Lemziakov }
\affiliation{Pico group, QTF Centre of Excellence, Department of Applied Physics,
Aalto University School of Science, P.O. Box 13500, 00076 Aalto, Finland}

\author{Dmitrii Lvov}
\affiliation{Pico group, QTF Centre of Excellence, Department of Applied Physics,
Aalto University School of Science, P.O. Box 13500, 00076 Aalto, Finland}

\author{Hiroto Mukai}
\affiliation{Department of Physics, Tokyo University of Science, 1--3 Kagurazaka, Shinjuku, Tokyo 162--0825, Japan}
\affiliation{RIKEN Center for Quantum Computing (RQC), 2--1 Hirosawa, Wako, Saitama 351--0198, Japan}

\author{Akiyoshi Tomonaga}
\affiliation{Department of Physics, Tokyo University of Science, 1--3 Kagurazaka, Shinjuku, Tokyo 162--0825, Japan}
\affiliation{RIKEN Center for Quantum Computing (RQC), 2--1 Hirosawa, Wako, Saitama 351--0198, Japan}

\author{Shintaro Takada}
\affiliation{National Institute of Advanced Industrial Science and Technology (AIST), National Metrology Institute of Japan (NMIJ), 1-1-1 Umezono, Tsukuba, Ibaraki 305-8563, Japan}

\author{Yuma Okazaki}
\affiliation{National Institute of Advanced Industrial Science and Technology (AIST), National Metrology Institute of Japan (NMIJ), 1-1-1 Umezono, Tsukuba, Ibaraki 305-8563, Japan}

\author{Nobu-Hisa Kaneko}
\affiliation{National Institute of Advanced Industrial Science and Technology (AIST), National Metrology Institute of Japan (NMIJ), 1-1-1 Umezono, Tsukuba, Ibaraki 305-8563, Japan}

\author{Jukka Pekola}
\affiliation{Pico group, QTF Centre of Excellence, Department of Applied Physics,
Aalto University School of Science, P.O. Box 13500, 00076 Aalto, Finland}

\author{Jaw-Shen Tsai}
\affiliation{Department of Physics, Tokyo University of Science, 1--3 Kagurazaka, Shinjuku, Tokyo 162--0825, Japan}
\affiliation{RIKEN Center for Quantum Computing (RQC), 2--1 Hirosawa, Wako, Saitama 351--0198, Japan}

\date{\today}

\begin{abstract}
Recent advancements in circuit quantum electrodynamics have enabled precise manipulation and detection of the single energy quantum in quantum systems. A quantum circuit refrigerator (QCR) is capable of electrically cooling the excited population of quantum systems, such as superconducting resonators and qubits, through photon-assisted tunneling of quasi-particles
within a \Erase{hybrid} superconductor-insulator-normal metal junction. In this study, we demonstrated \sn{instantaneous QCR in the quantum regime}. We performed the time-resolved measurement of the QCR-induced cooling of photon number inside the superconducting resonator by harnessing a qubit as a photon detector. From the enhanced photon loss rate of the resonator estimated from the amount of the AC Stark shift, the QCR was shown to have
a cooling power of approximately 300 aW.\Erase{the experiments with this scheme clearly demonstrated that the number of photons inside the resonator could be lowered compared to that in the state of thermal equilibrium at various temperatures.} Furthermore, \sn{even below the single energy quantum, the QCR can reduce the number of photons inside the resonator with 100 ns pulse from thermal equilibrium.} Numerical calculations based on the Lindblad master equation successfully reproduced thees experimental results. 

\end{abstract}

\maketitle


Recently, there has been increasing interest in quantum technologies, such as quantum computation\cite{Arute2019, McArdle2020, Emani2021, Herman2023}, quantum communication\cite{Zhang2022, Nadlinger2022, Joshi2020}, and quantum metrology\cite{Takamoto2020, Hosten2016, Glenn2018}. These quantum-based technologies offer novel functionalities and sensitivities that are not attainable using classical technologies. However, numerous studies have shown that maintaining quantum states can be challenging because of thermal disturbances\cite{Krinner2019, Paik2011}  and external environmental noise\cite{Schlosshauer2005}.


The superconductor/insulator/normal-metal (SIN) (and its series connection SINIS) junction has emerged as one of the most powerful methods for cooling nano-fabricated electrical circuits\cite{Nahum1994, Leivo1996, Juha2012}. Previous studies have demonstrated that the electron temperature of a normal metal in a voltage-biased junction can be effectively reduced by the elastic tunneling of higher-energy electrons from the normal metal to the superconductor. Furthermore, because a current-biased junction also acts as a thermometer, the \Erase{hybrid} SIN junction provides opportunities to explore the thermodynamics \sn{not only} in nanoscale electrical circuits\cite{Giazotto2006} \sn{, but also in circuit quantum electrodynamics\cite{Ronzani2018, Senior2020, Gubaydullin2022}}. 

\Erase{Furthermore, utilizing this hybrid SIN junction in}
\Erase{ circuit quantum electrodynamics has attracted} 
\Erase{significant attention for mitigating thermal disturbances} 
\Erase{and for controlling the heat through a quantum system} 
\Erase{whose energy structure is well designed [18-20].}



Recently, it has been demonstrated that the inelastic tunneling process of the SIN junction plays an important role in absorbing energy from quantum circuits\cite{Tan2017, Mörstedt2022}. Under certain bias conditions\Erase{ of the SIN}, quasiparticles inside the superconducting electrodes can tunnel through the insulator via photon absorption. By attaching a junction to a quantum system, this process can absorb energy in the form of photons from the quantum system. This\Erase{ photon absorption with the SIN junction} is called quantum circuit refrigerator (QCR), and at present, accelerating initialization of a superconducting qubit\cite{Serviuk2022, Yoshioka2023}, cooling photon population inside the resonator\cite{Masuda2018}, and the generation of incoherent microwave photons\cite{Hyyppa2019} have been demonstrated\Erase{ with the QCR}. 

\sn{The next major experimental achievement of this new cooling technique is higher fidelity qubit initialization in a shorter time. One of the most promising protocols is the reset scheme using a three-level\Erase{ transmon artificial atom } \sn{system} coupled to a resonator\cite{Magnard2018}, whose relaxation rate is enhanced by the QCR\cite{Yoshioka2023}. In these reset schemes, the qubit excited state is converted to the photon state in the cavity. So, to confirm the higher fidelity and the faster initialization with the QCR, we should check not only the qubit state but also the resonator photon state immediately after instantaneous photon absorption with accuracy much better than a single energy quantum\cite{Zhou2021, MacClure2016}.}\sn{However, in previous experiments, the cooling of photon numbers with the QCR was assessed using commercially available room temperature apparatus\cite{Masuda2018, Sevriuk2019}. These measurements were notably influenced by substantial offset noise ($\sim$ 3 K), leading to a reduced signal-to-noise ratio. Consequently, accurately estimating the instantaneous reduction in photon numbers below a single energy quantum presents a significant challenge.}

\begin{figure}[t]
 \begin{center}
  \includegraphics[width=9cm]{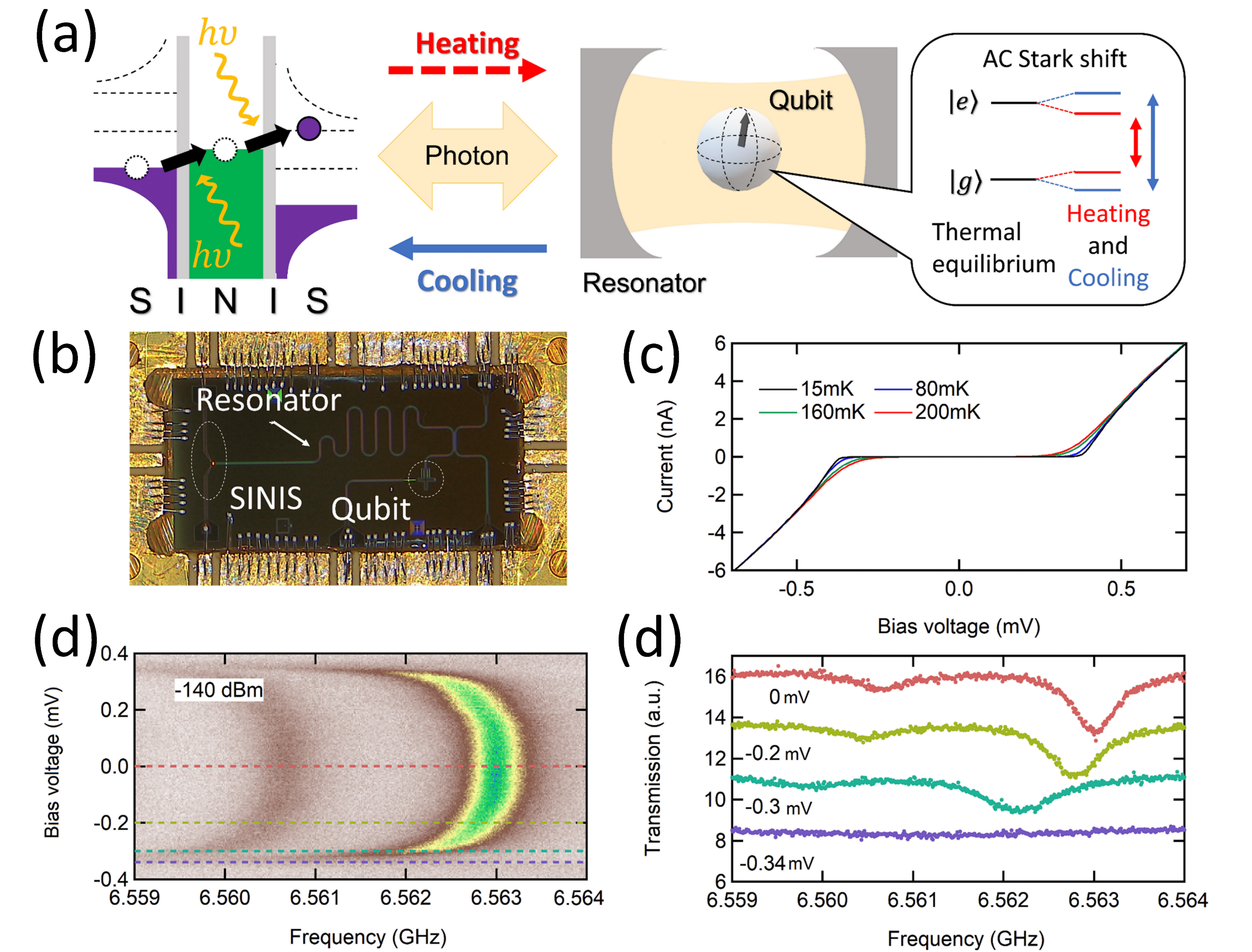}
  \caption{(a) Schematic picture of the detection of the QCR-induced photon cooling (heating) with the AC-Stark shift of the qubit. (b) The optical microscope image of the whole sample on a printed circuit board. (c) The current-voltage ($I-V$) characteristic of the SINIS junction at various temperatures of dilution refrigerator (Black: 15 mK, blue: 80 mK, light green: 160 mK, and red: 200 mK) (d) Plot of the bias dependence of transmission spectrum of the superconducting resonator which is coupled to the transmon and SINIS. The probed power is fixed at -140 dBm. (e) The line profiles of the image plot (d) at the various SINIS bias conditions (Pink: 0 mV, Yellow: -0.2 mV, Green: -0.3 mV, Purple: -0.34 mV). }
  \label{}
 \end{center}
\end{figure}

\sn{Here,}\Erase{Hybrid quantum systems provide promising} \Erase{architectures for the highly sensitive detection of various}\Erase{elementary excitations in quantum systems[28-30].}\Erase{In this study,} we demonstrate the precise measurement of the \sn{instantaneous 
quantum circuit refrigeration}\Erase{QCR-induced cooling} of the photon \Erase{population} inside the resonator \sn{below a single energy quantum.}\Erase{ with a superconducting qubit.} We performed a time-resolved measurement of cooling inside a superconducting resonator, probed by the AC-Stark shift of the qubit.\Erase{ In the dispersive limit, the qubit resonance is shifted owing to the photon fields inside the resonator, and such frequency shifts are proportional to the photon number $n_\mathrm{photon}$ inside.} Through spectroscopic measurement of the qubit with a microwave pulse after manipulating the photon population of the resonator with the QCR, we measured the instantaneous photon number inside the resonator.\Erase{ Despite previous studies probing photon cooling with the room temperature commercial apparatus\cite{Masuda2018}} \sn{Compared with the previous experiment}, our measurement involved a qubit that was directly coupled to the resonator, acting as a photon detector. This onsite measurement scheme is more sensitive to the photon number inside the resonator\Erase{ and is hardly affected by a significant amount of offset noise ($\sim 4$ K) from the HEMT amplifier. As a result}, and we successfully demonstrated that the QCR could cool the photon population of the resonator \sn{even} below \sn{the single energy quantum with 100 ns pulse from thermal equilibrium.}\Erase{ one at thermal equilibrium up to 200 mK}. \Erase{Moreover, \sn{we show the number of photon absorption} \Erase{cooling power increases as the temperature increases}, which is well reproduced by the numerical simulation using the Lindblad master equation.}

In this study, we utilized a transmon qubit, which was capacitively coupled to a half-wavelength coplanar waveguide (CPW) resonator on one end. The SINIS structure was placed on the other end of the resonator for the QCR. The CPW resonator was capacitively coupled to the normal metal of the SINIS using an interdigital capacitor. To read the qubit state, the transmission line was \sn{also} capacitively coupled to the CPW resonator. Figure 1 (c) shows the current-voltage ($I$-$V$) characteristics of SINIS at different temperatures. The current was suppressed owing to the superconducting gap $\mathit{\Delta}$ at the bias voltages $\left|V_\mathrm{SD}\right| < 2\mathit{\Delta}/e$ (where $e$ represents the elementary charge); however, it started to flow above the bias voltages $\left|V_\mathrm{SD}\right| > 2\mathit{\Delta}/e$ owing to the sequential quasiparticle tunneling from the source to the drain superconducting-electrode. The elevated temperature induced the additional quasiparticles inside the superconductor leads and changed the Fermi-Dirac distribution of the electron inside the normal metal. This increase in temperature caused the additional current around the bias voltage  $\left|V_\mathrm{SD}\right| \sim 2\mathit{\Delta}/e$. From the numerical fitting of the $I$-$V$  characteristic, the tunnel resistance is $R_\mathrm{T}$ = 44 k$\Omega$ and the Dynes parameter $\gamma_{\mathrm{D}} = 1.8 \times 10^{-4}$ .

Next, we characterized the superconducting resonator having a transmon and a SINIS at each end. In the dispersive limit, where the coupling strength $g/2\pi$ between the qubit and the resonator is much smaller than their frequency difference $\delta_\mathrm{d}/2\pi = (\omega_\mathrm{r}/2\pi - \omega_\mathrm{ge}/2\pi)$, the Jaynes-Cummings Hamiltonian of the hybrid system is described by
\begin{equation}
\label{JC Hamiltonian}
{\hat{H}/\hbar \approx \omega_\mathrm{r}(\hat{a}^\dagger\hat{a}+\frac{1}{2}) + \frac{1}{2}(\omega_\mathrm{ge} + 2\chi \hat{a}^\dagger\hat{a} + \chi)\sigma_{z}},
\end{equation}
where $\hat{a}^\dagger$, $\hat{a}$ are photon creation and annihilation operator, and $\sigma_{z}$ is Pauli matrix for the qubit. In this limit, the qubit and resonator cannot exchange energy; however, the frequency of the qubit and resonator is shifted owing to renormalization by the dispersive interaction. The resonant frequency of the resonator deviates from the bare resonance by the qubit state-dependent shift of $ \pm \chi/2\pi$. Figures 1 (d) and (e) show the results of the transmission spectrum of the transmission line capacitively coupled to the resonator. When the bias voltage $\left|V_\mathrm{SD}\right|$ of the SINIS is much smaller than the value $2\mathit{\Delta}/e$, the two different resonance dips are observed owing to the thermal population of the qubit. Based on this measurement, we can estimate the dispersive interaction strength $ \chi/2\pi \sim$ 1.25 MHz. \sn{(Also, from another measurement, we estimated the bare frequency of the resonator $\omega_\mathrm{r}/2\pi$ = 6.538 GHz and the bare resonator relaxation rate $\kappa_\mathrm{r}/2\pi$ = 3.8 MHz.)} As the bias voltage became close to the value of $2\mathit{\Delta}/e$, the enhanced photon loss owing to the photon-assisted tunneling (PAT) inside the SINIS widened the width of the resonance, and the frequency shift appeared owing to the broadband Lamb shift, as shown in previous studies\cite{Silveri2019}.


\begin{figure}[t]
 \begin{center}
  \includegraphics[width=8.4cm]{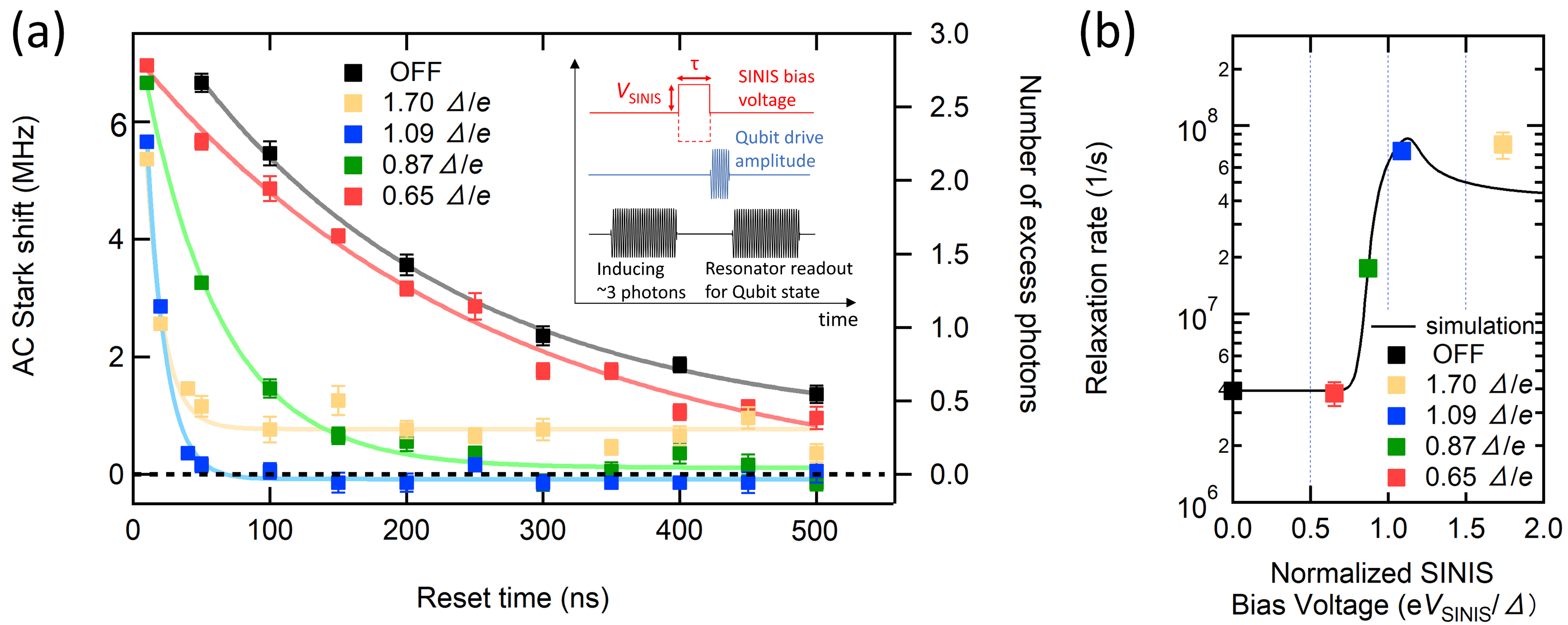}
  \caption{Time-resolved measurement of cooling of photons inside the resonator (a) The relaxation of the injected photons at various bias voltages to the SINIS  ($V_\mathrm{SINIS}$ = OFF (black squares), $0.65\times\mathit{\Delta}/e$ (red squares),  $0.87\times\mathit{\Delta}/e$ (green squares),  $1.09\times\mathit{\Delta}/e$(blue squares),  $1.70\times\mathit{\Delta}/e$ (yellow squares)). The black dotted line shows the frequency at thermal equilibrium. (inset) Pulse sequences for the measurements. \sn{Injected photon in the SINIS-OFF sate is slightly (20 \%) changed from the other SINIS-ON states.}(b) The relaxation rate at each bias voltage of SINIS. The relaxation rate is obtained by numerical fitting with the exponential (Colored solid lines in Fig. 2(a)). The error bars in Figures (a) and (b) indicate the error bars of the numerical fitting. The black curve indicates the result of the numerical calculation with the parameters of TABLE I in Supplemental Materials.}
  \label{}
 \end{center}
\end{figure}

To probe the QCR-induced absorption and emission of photons inside the resonator, we used a transmon qubit. In the qubit case, the frequency shift is the sum of Lamb shift and AC Stark shift which is proportional to the intra-resonator photon number $\langle n \rangle = \hat{a}^\dagger\hat{a}$. \sn{We conducted proper calibration of the photon number inside the cavity and confirmed that the resonant frequency of the qubit was proportionally shifted by the AC Stark effect from the photon fields inside the superconducting resonator (See Supplemental Information).}\Erase{Figures 1 (f) and (g) show the results of the qubit spectroscopy with the conventional dispersive readout after applying a 3 {\textmu}s resonator-drive pulse at various amplitudes of the resonator-drive microwave. Because the resonator’s internal energy (photon number) is proportional to the square of the amplitude of the resonator-drive microwave, the frequency of the qubit should be quadratically changed against the drive amplitude. The experimental data clearly showed that the qubit resonance depends on the photon number inside the resonator, as expected from AC stark shift. Here, the width of the resonance peak of the qubit even at the lowest resonator-drive (0.1 V) is relatively wider compared to the value estimated from the qubit coherence time $T_{1} \sim$ 9 {\textmu}s. This is because} \sn{Here,} we use a relatively strong pulse for the qubit spectroscopy to ensure that the $\pi$-pulse of the qubit becomes $\sim$ 100 ns, which is shorter than the relaxation time of the resonator $\sim$ 260 ns estimated from the relaxation rate at the off state of the QCR. \sn{Due to this relaxation, the measured photon number is about 68 \% of the one immediately after the cooling or heating completion.} Consequently, this approach enables us to detect the average of the photon number inside the resonator before the induced photons are completely depleted.


After confirming that the resonant frequency of the qubit was shifted by the AC Stark effect from the photon fields inside the superconducting resonator, we performed time-resolved measurements of the QCR-induced photon cooling. The inset of Fig. 2 (a) shows the pulse sequence. First, we injected a 5 {\textmu}s burst signal to the resonator at the \sn{dressed resonant} frequency \sn{$\tilde{\omega_\textrm{r}}$}, whose energy corresponds to $\sim$ 3 photons. Next, the SINIS was turned on during certain times with the square pulse at the various amplitudes ($V_\mathrm{SINIS}$ = OFF, $0.65\times\mathit{\Delta}/e,  0.87\times\mathit{\Delta}/e,  1.09\times\mathit{\Delta}/e,  1.70\times\mathit{\Delta}/e $) to cool down the photons inside the resonator. Note that, in our experiment, the bias for the QCR is applied at both the source and drain sides of the leads with square pulses\sn{, where voltages of opposite polarity but equal magnitude are applied}. Subsequently, we performed spectroscopy of the qubit with a 100 ns pulse and estimated the number of photons immediately after cooling. Here the qubit relaxation and photon relaxation after the pulse for the qubit-spectroscopy should not play important roles in determining the shifted resonant frequency of the qubit. Figure 2 (a) shows the time-resolved measurement of the excess photon number from the thermal equilibrium value (black dotted line in Fig. 2 (a)). When the QCR is in the OFF state, the induced $\sim 3$ photons inside the resonator decay at the rate of bare photon loss (black squares in Fig. 2 (a)). When the pulse amplitude for the QCR approached the value $V_\mathrm{SINIS} \sim \mathit{\Delta}/e $ (red, green, and blue squares in Fig. 2 (a)), the induced photons were reduced at the enhanced relaxation rate owing to the QCR. Figure 2 (b) shows the estimated photon relaxation rate of the resonator. The enhanced relaxation rate with the QCR was approximately 20 times the photon loss at $V_\mathrm{SINIS} = 0$ (OFF). Moreover, when the pulse height was fixed above the value $\mathit{\Delta}/e$ (yellow squares in Fig. 2 (a)), the decay rate of the photons hardly changed \sn{compared to the one at $V_\mathrm{SINIS}=\Delta /e$}; however, \sn{$\sim$ 0.4} \Erase{a few} photons remained inside the cavity owing to the incoherent photon emission from the QCR\cite{Masuda2018} \sn{even at long reset time}. From the relaxation of the photon number \sn{$\sim$ 74 photons/\textmu s at the value of $V_\mathrm{SINIS} = \mathit{\Delta}/e $}, we roughly estimated the cooling power of the QCR to be $P_\mathrm{SINIS} \equiv \hbar\sn{\tilde{\omega_\textrm{r}}}\times 7.4 \times 10^{7}$/s $\sim 320$ aW at the bias-voltage $V_\mathrm{SINIS}\sim\mathit{\Delta}/e$.

\begin{figure}[t]
 \begin{center}
  \includegraphics[width=8.5cm]{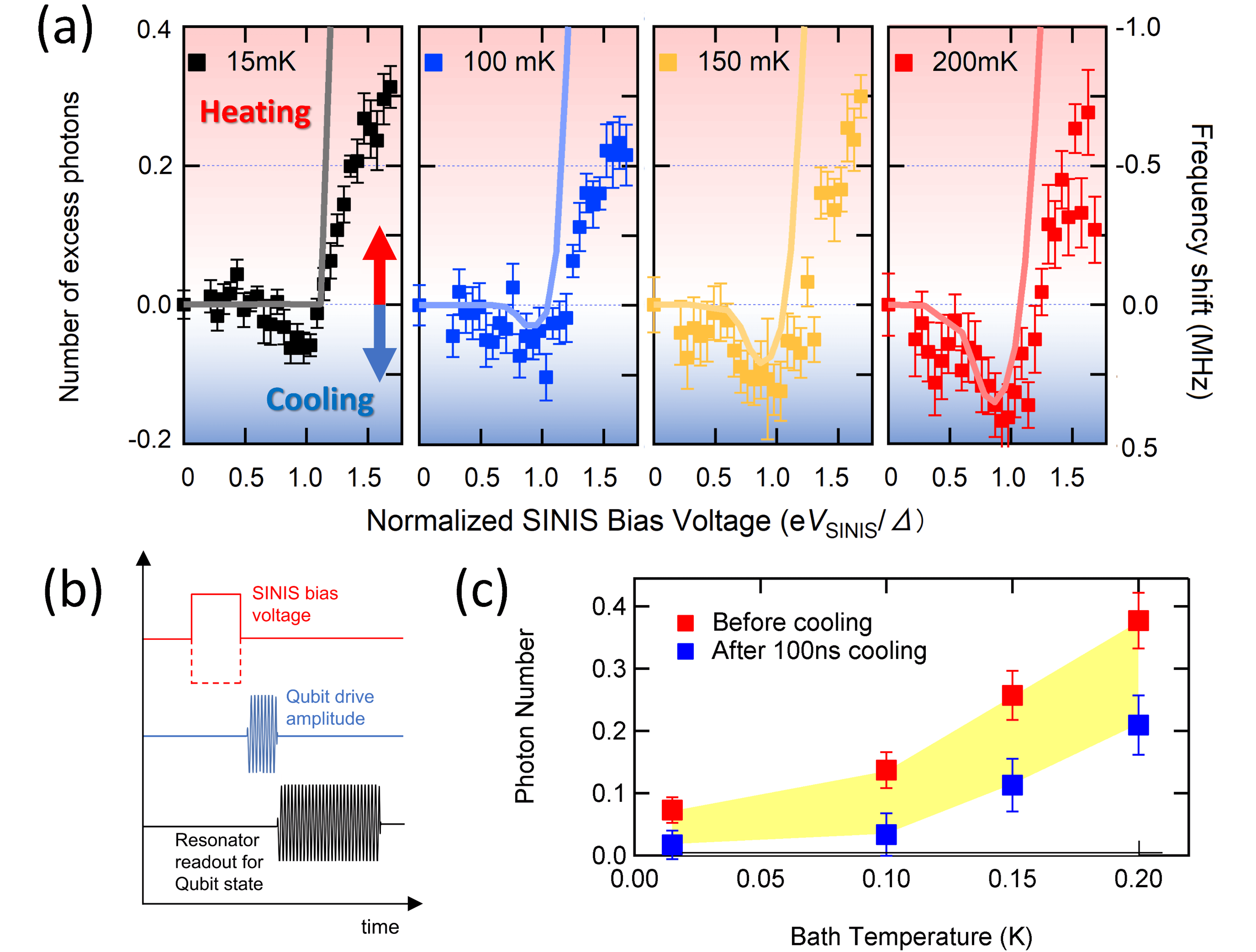}
  \caption{QCR-induced cooling of the photon population below the thermal equilibrium value. Experimental \sn{(colored squares)} and numerically calculated \sn{(colored solid lines)} frequency shift of the qubit from \sn{the} one at the thermal equilibrium states (right vertical axis) at various temperatures (black\Erase{ squares}: 15 mK, blue\Erase{ squares}: 100 mK, yellow\Erase{ squares}: 150 mK, and red\Erase{ squares}: 200 mK) and the number of excess photons (left vertical axis) estimated from the strength of the dispersive interaction $\chi$. The horizontal axis shows the pulse height of the bias voltage for the SINIS, which is normalized by the superconducting gap $\mathit{\Delta}$. (\sn{b}) Pulse sequences for the measurement. In this experiment, QCR-induced photon cooling started from the thermal equilibrium value without the injection of photons. (\sn{c}) \sn{Absolute photon number in the superconducting resonator before (red) and after (blue) cooling at the pulse height for the SINIS $V_\textrm{SINIS}= \mathit{\Delta}/e$.} \Erase{The photon absorption against the bath temperature of the dilution refrigerator when the pulse height for the SINIS is fixed at $V_\textrm{SINIS}= \mathit{\Delta}/e$. The red squares are experimental data. The \Erase{blue and }black squares are the results of numerical calculations.} \Erase{with different electron temperatures (blue squares: $T_{N} = T_\mathrm{Bath}$ - 20 mK) and black squares ($T_\mathrm{N} = T_\mathrm{Bath}$))}}
  \label{}
 \end{center}
\end{figure}

In addition to the enhanced relaxation rate, the saturation value of the excess photon number at $V_\mathrm{SINIS} \sim \mathit{\Delta}/e$ (blue squares in Fig. 2 (a)) became slightly ($\sim$ 0.06 photon) negative (less than the thermal equilibrium values (dotted line in Fig. 2 (a)) owing to the photon absorption with the QCR. \sn{Next,} to investigate this photon cooling below the thermal equilibrium values more precisely, we measured the bias-voltage dependence of the cooling at various \sn{bath} temperatures \sn{$T_\mathrm{Bath}$ (temperature of the mixing chamber plate in the dilution refrigerator)} with \sn{four times} long\sn{er} integration time \sn{compared to the one in Fig. 2 (a)}. In this experiment, we started cooling from the thermal equilibrium without injecting photons before cooling and fixed the cooling time to 100 ns, which is almost 7 times longer than the photon relaxation time at $V_\mathrm{SINIS} \sim \mathit{\Delta}/e$. Figure 3 (a) shows the SINIS bias pulse-height dependence of the number of generated photons compared to the thermal equilibrium value ($V_\mathrm{SINIS}$ = 0 V) at each temperature. At all temperatures, the generated photon number can be clearly negative around the $V_\mathrm{SINIS} \sim \mathit{\Delta}/e$ by the QCR, namely the photon population becomes smaller compared to the thermal equilibrium values. On the other hand, when the pulse amplitude becomes larger than $V_\mathrm{SINIS} \sim \mathit{\Delta}/e$, the number of generated photons gradually increased due to the photon emission of the QCR. Furthermore, the amount of photon absorption clearly increased when the bath temperature \sn{$T_{\mathrm{Bath}}$} became higher. \Erase{Here, the energy stored in the resonator is the sum of photon absorption or emission by the SINIS,\Erase{ energy loss to the substrate through phonons,} the influence of the transmission line coupled to the resonator, and internal photon loss from the resonator itself. Of these, the time scales for\Erase{ ’energy loss to the substrate through phonons’}\Erase{ and} internal photon loss of the resonator \Erase{itself }must be sufficiently longer than the time required for spectroscopy of the qubit (100 ns); thus, they can be neglected.}

To support these experimental observations theoretically, we numerically estimated the number of photons inside the resonator after cooling. We consider a harmonic oscillator coupled to two different thermal baths: the transmission line and the QCR. From the coupling strengths ($\gamma_\mathrm{tr}$, $\gamma_\mathrm{\sn{N}}$) and the mean thermal occupations ($\mathcal{N}_\mathrm{tr}$, $\mathcal{N}_\mathrm{\sn{N}}$) of the transmission line and the normal metal, we can calculate the photon number inside the resonator with the Lindblad master equation as follows:

\begin{equation}
\label{Lindblad}
\begin{split}
\Dot{{\hat\rho}} = -\frac{i}{\hbar}[\hat{H_\mathrm{r}}, \hat\rho] & + \{ \gamma_\mathrm{\sn{N}}(\mathcal{N}_\mathrm{\sn{N}} +1) + \gamma_\mathrm{tr}(\mathcal{N}_\mathrm{tr} +1)\}\mathcal{D}(\hat{a})\hat{\rho} \\& + (\gamma_\mathrm{\sn{N}}\mathcal{N}_\mathrm{\sn{N}} + \gamma_\mathrm{tr}\mathcal{N}_\mathrm{tr})\mathcal{D}(\hat{a}^{\dagger})\hat{\rho} 
\end{split}
\end{equation}
where $\rho$ is the density matrix of the resonator, $\hat{H_\mathrm{r}}$ is the Hamiltonian of the resonator, and the $\mathcal{D}(\hat{A})\hat{\rho} = \hat{A}\rho\hat{A}^{\dagger} - \frac{1}{2} \{\sn{\hat{A}^{\dagger}\hat{A}}, \rho \}$. Coupling strength $\gamma_\mathrm{tr}$, associated with the transmission line thermal bath, can be determined from the resonator quality factor \sn{at zero SINIS bias voltage.} The mean thermal occupation $\mathcal{N}_\mathrm{tr}$ is the Bose-Einstein distribution for the resonator frequency and phonon temperature at the sample position. \sn{Here, we ignore the internal photon loss of the resonator itself because its relaxation time is much longer than the time for the spectroscopy (100 ns).} 


Utilizing the model for the influence of PAT on the resonator decay rate and photon number from \cite{Silveri2017}, the expressions for the decay rate and photon number are as follows:
\begin{align} \label{decay rate}
    {\gamma}_\mathrm{\sn{N}} = \gamma_\mathrm{\sn{N}}^0 \frac{\pi}{\sn{\tilde{\omega_\mathrm{r}}}}
    \sum_{\tau,\ell=\pm1} \mathcal{F}(\tau eV + \hbar\sn{\tilde{\omega_\mathrm{r}}}\ell-E_\mathrm{N})\,, 
\end{align}

\begin{align} \label{decay rate}
    \gamma_\mathrm{\sn{N}}^0 = 2 \frac{C_\mathrm{C}^2}{(C_\mathrm{N} + C_\mathrm{C})^2}\frac{Z_\mathrm{r}}{R_\mathrm{T}}\sn{\tilde{\omega_\mathrm{r}}}\,,
\end{align}

\begin{align} \label{photon number}
    \mathcal{N}_\mathrm{\sn{N}} = (\frac{\sum_{\tau=\pm1} \mathcal{F}(\tau eV + \hbar\sn{\tilde{\omega_\mathrm{r}}}-E_\mathrm{N})}{\sum_{\tau=\pm1} \mathcal{F}(\tau eV - \hbar\sn{\tilde{\omega_\mathrm{r}}}-E_\mathrm{N})}-1)^{-1}\,, 
\end{align}
where $\mathcal{F}$ is \Erase{a function of the state distribution}\sn{a normalized rate of quasiparticle tunneling} (please see Supplement), $Z_r$ is the impedance of the resonator, $C_\mathrm{C}$ is the capacitance between the normal metal island and the resonator, $C_\mathrm{N}$ is the capacitance of the normal metal to the ground, \sn{$l$ and $\tau$ are the signs associated with the absorption and emission processes of photons and the polarity of the bias respectively, $E_{N}$ is the change in the charging energy of the normal metal island due to tunneling, $f_{\mathrm{i}}$($i$ = S,N) is the Fermi-Dirac distribution function, and $n_{\mathrm{S}}$ is the density of state of the superconductor}. The \Erase{distribution} function $\mathcal{F}$ depends on the normal metal electron temperature $T_\mathrm{N}$, which also depends on the SINIS voltage $V$ owing to electron cooling\cite{Nahum1994, Giazotto2006}. \sn{Here, we assume that the temperature of the superconducting leads are equal to the bath temperature.} \sn{Also, to simplify the calculation, the broadband Lamb shift of the dressed resonator frequency \sn{$\tilde{\omega_\mathrm{r}}$}, the electron cooling and heating were not taken into account.}


Despite the master equation (\ref{Lindblad}) having a simple steady-state analytical solution, to consider the decay of the photon number into the resonator after the pulse on the SINIS, we solved the time-dependent master equation numerically. To perform it, we used a steady-state solution at zero voltage on the SINIS as an initial condition. Starting from it, we calculated the time-dependent density matrix for 100 ns during the square voltage pulse on the SINIS and the next 100 ns immediately after the pulse. The last 100 ns corresponded to the time required for qubit excitation during the measurement protocol. The final result for the calculated photon number in the cavity was obtained by averaging the expectation value of $\hat{a}^{\dagger}\hat{a}$ during the last 100 ns. 


\sn{The results of numerical calculation at various temperatures $T_\mathrm{Bath}$ are depicted in the colored solid lines in Fig. 3(a)}\Erase{Figure 3(b) show the results of numerical calculations}. \sn{With respect to the photon absorption by the QCR,} the numerical calculations\Erase{ of photon absorption by the QCR} have almost quantitatively reproduced the experimental results concerning the bias dependence \sn{except for the lowest bath temperature $T_\mathrm{bath} =$ 15 mK. In the theoretical model, it is assumed that the electron temperature of the device is the same as the bath temperature of 15 mK. This assumption leads to almost no photons in the resonator, hence, theoretical calculations show no photon absorption at $V_\mathrm{SINIS} = \Delta/e$. However, in the experiment, the electron temperature of the device was not thermalized at the bath temperature $T_{Bath}$, and the electron temperature of the device estimated from Rabi population measurement\cite{Jin 2015} at $T_\mathrm{Bath}$ = 15 mK is $\sim$ 120 mK (see Supplement).
Therefore, we observed the some amount of photon absorption even at $T_\mathrm{Bath}$ = 15 mK. This kind of deviation between electron temperature and the bath temperature has been well-known for a long time\cite{Wellstood1994}. Related to the temperature dependence of the photon absorption, this numerical calculation reproduces the behavior of the increased absorption of photons at $V_\mathrm{SINIS} =\mathit{\Delta}/e$ when the bath temperature becomes higher.} 

\Erase{\sn{This deviation that the electron temperature deviates from the bath temperature (especially below 100 mK) has often been reported in cryogenic experiments. The electron temperature estimated from two different method clearly shows The electron temperature estimated from the $I-V$ characteristic of SINIS ($T_{N}$ = 70 mK)and the finite population of excited states in our qubits ($\sim 12 \%$ which equals to the population at 120 mK) at base temperature ($T_{Bath}$ = 15 mK), indicating that the actual electron temperature of the device exceeds the bath temperature. Moreover, the theory also accurately replicates the behavior of the increased absorption of photons at $V_\mathrm{SINIS} =\mathit{\Delta}/e$ when the bath temperature becomes higher as shown in Fig. 3 (c).}}

\Erase{\sn{Also in this photon absorption, we may observe a slight increase in photon number around $V_\mathrm{SINIS} = \Delta/3e$. In the theory of the QCR\cite{Silveri2017}, the Dynes parameter induces the increase of photons, which take a maximum at a third of the superconducting gap even when a bias voltage is below the superconducting gap. Due to the large error bar, we cannot confirm that the slight increase in photon emission corresponds to this phenomenon. A large cavity pull system may help to address this issue.}}

\sn{On the other hand,} regarding the photon emission when $V_\mathrm{SINIS}$ becomes large, the experimental results display less emission than the theoretical prediction. \Erase{Also in this case, the electron cooling should be attributed to the deviation, however, further investigation is necessary}\sn{ There are several factors that may contribute to the discrepancy between the theoretical calculations and the experimental results. For example, we did not take into account the electron heating and cooling, the difference in the temperature between the normal metal and superconducting lead, the frequency shift of the resonator due to the broadband Lamb shift, and imperfections in the pulse shape. To fully understand the reasons for the reduced photon emission, further theoretical and experimental investigations are needed (see Supplement).}

\Erase{This is because the electron temperature of the device deviates from the bath temperature, specifically, the temperature of the thermometer on the mixing chamber plate. Assuming the electronic temperature of the devices themselves is at the dilution refrigerator's temperature of 15 mK, the theoretically estimated number of photons inside the resonator is almost zero, which leads to no observed photon absorption. However, experimentally, it is often observed that the electron temperature of the device deviates from the bath temperature $T_\mathrm{Bath}$. If the electron temperature of the device reaches around 100 mK, the experimental results align well with the calculations. This elevated electron temperature is also observed in other measurements, such as the $I-V$ measurement of SINIS and the large excited population of the qubit in our experiment.}

\Erase{Figure 3 (d) shows photon absorption at various temperatures when we fixed $V_\mathrm{SINIS} = \mathit{\Delta}/e$. Both experimental data and the theoretical prediction clearly showed that photon absorption becomes larger as the bath temperature becomes higher.  Here, the experimental result of photon absorption is relatively larger than the theoretical values. One reason for this deviation can be attributed to the fact that the temperature of the normal metal in the QCR deviated from the bath temperature due to the inadequate thermalization of the device. Actually, the electron temperature of the normal metal, obtained from the numerical fitting of the $I-V$ characteristic at the lowest bath temperature, is approximately 70 mK. Another reason for the relatively large photon absorption is the electron cooling of the normal metal. In fact, the maximum photon absorption becomes larger if the electron temperature of the normal metal is cooled down by the electron cooling as shown in the blue ($T_{N}$ = $T_\mathrm{Bath} - 20$ mK) and the black squares ($T_{N}$ = $T_\mathrm{Bath}$) of Fig. 3 (d). Regarding the photon emission when $V_\mathrm{SINIS} > \mathit{\Delta}/e$, the experimental results display less emission than the theoretical prediction. Also in this case, the electron cooling should be attributed to the deviation, however, further investigation is necessary. }

\sn{Finally, we estimated the absolute number of photons before and after cooling at each bath temperature (Fig. 3(c)). From the device temperature estimated from the Rabi population measurement at the lowest bath temperature ($T_\mathrm{Bath}$ =15 mK) and the qubit frequency shift, we determined the absolute number of photons before and after cooling. As a result, we showed that the QCR is effective even below the single energy quantum; namely, it is in the quantum regime.}

\Erase{In this study}\sn{Here}, we demonstrated \sn{fast and} precise measurement of QCR-induced photon cooling inside a superconducting resonator \sn{below a single energy quantum.}\Erase{by harnessing the superconducting qubit as a photon detector.} \sn{In this experiment with an integration time of approximately one minute, the fitting error bars for estimating the photon number are around 0.02 to 0.03 photons. Therefore, by increasing the integration time by a factor of 4 to 9, it is possible to adequately measure the photon number immediately after initialization with a photon resolution of less than 0.01 photons, which helps to confirm qubit initialization.} \sn{Thus, this study represents a crucial step towards the evaluation of faster and higher fidelity qubit initialization. Also, It paves the way for future research, such as the utilization and evaluation of quantum circuit refrigeration in the phonon systems. Also, with adjustments of the cavity pull\cite{Gambetta2006}\cite{Arto2023}, the distribution of photon occupancy in the resonator immediately after QCR-induced cooling can be accessed.}
\Erase{\sn{Also,}  owing to the relatively large frequency detuning between the qubit and the resonator, we could only determine the average number of photons. \Erase{However, w}\sn{W}ith adjustments to appropriate values\cite{Gambetta2006}, the distribution of photon occupancy in the resonator after QCR-induced cooling can be determined. Recently, a paper by Viitanen et al. appeared with a similar setup, and demonstrated measurement of the photon occupancy with the QCR utilizing a larger cavity pull system. }

\Erase{This study represents a crucial step towards the faster and higher fidelity qubit initialization, as well as the utilization and evaluation of the quantum circuit refrigeration in other quantum systems, such as phonon systems. It paves the way for future research developments in these areas.We observed \sn{instantaneous} photon absorption \sn{from initial thermal equilibrium state below the single energy quantum.}\Erase{ and its temperature dependence, which is consistent with the theoretical calculations.} }

We thank Prof. Yasuhiro Tokura, Prof. Mikko M\"ottonen, and Dr. Tsuyoshi Yamamoto for their fruitful discussion. S.N. acknowledges the financial support by JSPS KAKENHI Grant Number 20KK0335 and 20H02561.



\begin{thebibliography}{99}

\bibitem{Arute2019} F. Arute, K. Arya, R. Babbush, et al., Quantum supremacy using a programmable superconducting processor. Nature 574, 505 (2019).
\bibitem{McArdle2020} S. McArdle, S. Endo, A. Aspuru-Guzik, S. C. Benjamin, and X. Yuan,X, Quantum computational chemistry, Rev. Mod. Phys. 92, 015003 (2020).
\bibitem{Emani2021} P. S. Emani, J. Warrell, A. Anticevic, et al. Quantum computing at the frontiers of biological sciences, Nature Methods 18, 701 (2021).
\bibitem{Herman2023} D. Herman, C. Googin, X. Liu, et al., Quantum computing for finance. Nature Rev. Phys. 5, 450 (2023). 
\bibitem{Zhang2022}W. Zhang, T. van Leent, K. Redeker, et al., A device-independent quantum key distribution system for distant users. Nature 607, 687 (2022). 
\bibitem{Nadlinger2022}D. P. Nadlinger, P. Drmota, B. C. Nichol, et al., Experimental quantum key distribution certified by Bell's theorem, Nature 607, 682 (2022). 
\bibitem{Joshi2020} S. K. Joshi, D. Aktas, S. Wengerowsky et. al., A trusted-node-free eight-user metropolitan quantum communication network, Science Advances 6, 36 (2020).
\bibitem{Takamoto2020}M. Takamoto, I. Ushijima, N. Ohmae, et al., Test of general relativity by a pair of transportable optical lattice clocks. Nature Photonics 14, 411 (2020).
\bibitem{Hosten2016} O. Hosten, N. Engelsen, R. Krishnakumar, et al., Measurement noise 100 times lower than the quantum-projection limit using entangled atoms. Nature 529, 505 (2016).
\bibitem{Glenn2018} D. Glenn, D. Bucher, J. Lee, et al., High-resolution magnetic resonance spectroscopy using a solid-state spin sensor. Nature 555, 351 (2018).
\bibitem{Krinner2019}S. Krinner, S. Storz, P. Kurpiers, et al., Engineering cryogenic setups for 100-qubit scale superconducting circuit systems, EPJ Quantum Technol. 6, 2 (2019).
\bibitem{Paik2011} H. Paik, D. I. Schuster, L. S. Bishop et. al., Observation of High Coherence in Josephson Junction Qubits Measured in a Three-Dimensional Circuit QED Architecture, Phys. Rev. Lett. 107, 240501 (2011).
\bibitem{Schlosshauer2005} M. Schlosshauer, Decoherence, the measurement problem, and interpretations of quantum mechanics, Rev. Mod. Phys. 76, 1267 (2005).
\bibitem{Nahum1994}M. Nahum, T. M. Eiles, John M. Martinis, Electronic microrefrigerator based on a normal‐insulatorsuperconductor tunnel junction, Appl. Phys. Lett. 65, 3123 (1994).
\bibitem{Leivo1996} M. M. Leivo, J. P. Pekola, and D. V. Averin, Efficient Peltier refrigeration by a pair of normal metal/insulator/superconductor junctions, Appl. Phys. Lett. 68, 1996 (1996).
\bibitem{Juha2012}Juha T. Muhonen, Matthias Meschke, Jukka P. Pekola, Micrometre-scale refrigerators, Reports on Progress in Physics  75, 046501 (2012).
\bibitem{Giazotto2006}F. Giazotto, T. T. Heikkilä, A. Luukanen, A. M. Savin, J. P. Pekola, Opportunities for mesoscopics in thermometry and refrigeration: Physics and applications, Rev. Mod. Phys. 78, 217 (2006)
\bibitem{Ronzani2018} A. Ronzani, B. Karimi, J. Senior, et al., Tunable photonic heat transport in a quantum heat valve. Nature Phys 14, 991 (2018).
\bibitem{Senior2020} J. Senior, A. Gubaydullin, B. Karimi, et al. Heat rectification via a superconducting artificial atom. Commun Phys 3, 40 (2020). 
\bibitem{Gubaydullin2022} A. Gubaydullin, G. Thomas, D. S. Golubev, et al., Photonic heat transport in three terminal superconducting circuit. Nat Commun 13, 1552 (2022).
\bibitem{Tan2017}K. Tan, M. Partanen, R. Lake, et al. Quantum-circuit refrigerator, Nature Communication 8, 15189 (2017).
\bibitem{Mörstedt2022}T. F. Mörstedt, A. Viitanen, V. Vadimov, et al., Recent Developments in Quantum-Circuit Refrigeration, Annalen der Physik 534, 2100543 (2022).
\bibitem{Serviuk2022}V. A. Sevriuk, W. Liu, J. Rönkkö et. al., Initial experimental results on a superconducting-qubit reset based on photon-assisted quasiparticle tunneling, Appl. Phys. Lett. 121, 234002 (2022). 
\bibitem{Yoshioka2023} Teruaki Yoshioka, Hiroto Mukai, Akiyoshi Tomonaga, Shintaro Takada, Yuma Okazaki, Nobu-Hisa Kaneko, Shuji Nakamura, and Jaw-Shen Tsai, Active initialization experiment of a superconducting qubit using a quantum circuit refrigerator, Phys. Rev. Applied 20, 044077 (2023).
\bibitem{Masuda2018}S. Masuda, K. Y. Tan, M. Partanen, et al., Observation of microwave absorption and emission from incoherent electron tunneling through a normal-metal–insulator–superconductor junction, Sci. Rep. 8, 3966 (2018).
\bibitem{Hyyppa2019} E. Hyyppä, M. Jenei, S. Masuda, et al., Calibration of cryogenic amplification chains using normal-metal–insulator–superconductor junctions, Appl. Phys. Lett. 114, 192603 (2019).
\bibitem{Magnard2018}P. Magnard, P. Kurpiers, B. Royer, T. Walter, J.-C. Besse, S. Gasparinetti, M. Pechal, J. Heinsoo, S. Storz, A. Blais, and A. Wallraff, Fast and Unconditional All-Microwave Reset of a Superconducting Qubit,
Phys. Rev. Lett. 121, 060502 (2018).
\bibitem{Zhou2021}Yu Zhou, Zhenxing Zhang, Zelong Yin, Sainan Huai, Xiu Gu, Xiong Xu, Jonathan Allcock, Fuming Liu, Guanglei Xi, Qiaonian Yu, Hualiang Zhang, Mengyu Zhang, Hekang Li, Xiaohui Song, Zhan Wang, Dongning Zheng, Shuoming An, Yarui Zheng and Shengyu Zhang "Rapid and unconditional parametric reset protocol for tunable superconducting qubits" Nature Communications 12, 5924 (2021) 
\bibitem{MacClure2016}D. T. McClure, Hanhee Paik, L. S. Bishop, M. Steffen, Jerry M. Chow, and Jay M. Gambetta, Rapid Driven Reset of a Qubit Readout Resonator, Phys. Rev. Applied 5, 011001(R) (2016).
\bibitem{Sevriuk2019}V. A. Sevriuk, K. Y. Tan, E. Hyyppä, M. Silveri, M. Partanen, M. Jenei, S. Masuda, J. Goetz, V. Vesterinen, L. Grönberg, M. Möttönen, Fast control of dissipation in a superconducting resonator, Appl. Phys. Lett. 115 082601 (2019).
\bibitem{Silveri2019} M. Silveri, S. Masuda, V. Sevriuk, et al. Broadband Lamb shift in an engineered quantum system. Nature Phys. 15, 533 (2019).
\bibitem{Silveri2017} M. Silveri, H. Grabert, S. Masuda, K. Y. Tan, and M. MöttönenTheory of quantum-circuit refrigeration by photon-assisted electron tunneling, Phys. Rev. B 96, 094524 (2017).
\bibitem{Jin 2015}X. Y. Jin, A. Kamal, A. P. Sears et al., Thermal and Residual Excited-State Population in a 3D Transmon Qubit, Phys. Rev. Lett. 114, 240501 (2015).
\bibitem{Wellstood1994}F. C. Wellstood, C. Urbina, and J. Clarke, Hot-electron effects in metals, Phys. Rev. B 49, 5942 (1994).
\bibitem{Gambetta2006} J. Gambetta, A. Blais, D. I. Schuster, A. Wallraff,/L. Frunzio, J. Majer, M. H. Devoret, S. M. Girvin, R. J. Schoelkopf, Qubit-photon interactions in a cavity: Measurement-induced dephasing and number splitting, Phys. Rev. A 74, 042318 (2006).
\bibitem{Arto2023}A. Viitanen, T. Mörstedt, W. S. Teixeira, et al., Quantum-circuit refrigeration of a superconducting microwave resonator well below a single quantum, arXiv:2308.00397



\end{thebibliography}
\end{document}


\title{\sn{Probing instantaneous quantum circuit refrigeration in the quantum regime}}

\author{Shuji Nakamura}\email{shuji.nakamura@aist.go.jp}
\affiliation{National Institute of Advanced Industrial Science and Technology (AIST), National Metrology Institute of Japan (NMIJ), 1-1-1 Umezono, Tsukuba, Ibaraki 305-8563, Japan}

\author{Teruaki Yoshioka}
\affiliation{Department of Physics, Tokyo University of Science, 1--3 Kagurazaka, Shinjuku, Tokyo 162--0825, Japan}
\affiliation{RIKEN Center for Quantum Computing (RQC), 2--1 Hirosawa, Wako, Saitama 351--0198, Japan}
\affiliation{National Institute of Advanced Industrial Science and Technology (AIST), National Metrology Institute of Japan (NMIJ), 1-1-1 Umezono, Tsukuba, Ibaraki 305-8563, Japan}

\author{Sergei Lemziakov }
\affiliation{Pico group, QTF Centre of Excellence, Department of Applied Physics,
Aalto University School of Science, P.O. Box 13500, 00076 Aalto, Finland}

\author{Dmitrii Lvov}
\affiliation{Pico group, QTF Centre of Excellence, Department of Applied Physics,
Aalto University School of Science, P.O. Box 13500, 00076 Aalto, Finland}

\author{Hiroto Mukai}
\affiliation{Department of Physics, Tokyo University of Science, 1--3 Kagurazaka, Shinjuku, Tokyo 162--0825, Japan}
\affiliation{RIKEN Center for Quantum Computing (RQC), 2--1 Hirosawa, Wako, Saitama 351--0198, Japan}

\author{Akiyoshi Tomonaga}
\affiliation{Department of Physics, Tokyo University of Science, 1--3 Kagurazaka, Shinjuku, Tokyo 162--0825, Japan}
\affiliation{RIKEN Center for Quantum Computing (RQC), 2--1 Hirosawa, Wako, Saitama 351--0198, Japan}

\author{Shintaro Takada}
\affiliation{National Institute of Advanced Industrial Science and Technology (AIST), National Metrology Institute of Japan (NMIJ), 1-1-1 Umezono, Tsukuba, Ibaraki 305-8563, Japan}

\author{Yuma Okazaki}
\affiliation{National Institute of Advanced Industrial Science and Technology (AIST), National Metrology Institute of Japan (NMIJ), 1-1-1 Umezono, Tsukuba, Ibaraki 305-8563, Japan}

\author{Nobu-Hisa Kaneko}
\affiliation{National Institute of Advanced Industrial Science and Technology (AIST), National Metrology Institute of Japan (NMIJ), 1-1-1 Umezono, Tsukuba, Ibaraki 305-8563, Japan}

\author{Jukka Pekola}
\affiliation{Pico group, QTF Centre of Excellence, Department of Applied Physics,
Aalto University School of Science, P.O. Box 13500, 00076 Aalto, Finland}

\author{Jaw-Shen Tsai}
\affiliation{Department of Physics, Tokyo University of Science, 1--3 Kagurazaka, Shinjuku, Tokyo 162--0825, Japan}
\affiliation{RIKEN Center for Quantum Computing (RQC), 2--1 Hirosawa, Wako, Saitama 351--0198, Japan}

\date{\today}

\maketitle

\section{Sample fabrication}

The transmission line, CPW resonator, and capacitor for the transmon qubit were patterned by reactive ion etching after sputtering 50 nm Nb on a Si wafer. Subsequently, we fabricated a Josephson junction for the transmon qubit using the shadow evaporation technique. First, we deposited 40 nm Al on top of the Nb. After oxidizing the deposited Al with oxygen to produce a thin insulation layer of aluminum oxide, we deposited 60 nm of aluminum. Finally, we fabricated a SINIS structure (40 nm Al with oxidation and 60 nm Cu) on the wafer using the shadow evaporation technique.

\begin{figure}[h]
 \begin{center}
  \includegraphics[width=15cm]{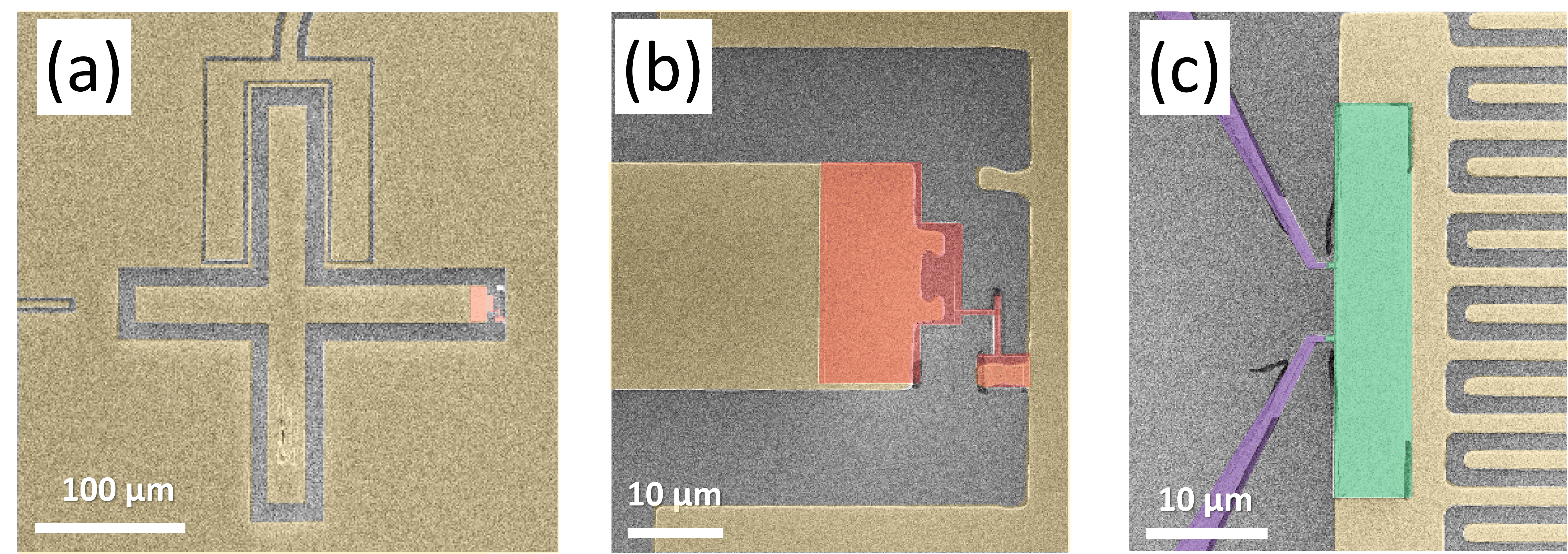}
  \caption{(a),(b),(c) The scanning electron microscope image of (a) the transmon qubit, (b) the Josephson junction and (c) the superconductor-insulator-normal-insulator-superconductor junction (SINIS) with the interdigital capacitor.}
  \label{}
 \end{center}
\end{figure}

\newpage

\section{Qubit and SINIS parameters}

\begin{table}[h]
  
  \label{table:sample}
  \centering
  \small
\begin{tabularx}{\textwidth}{Xcc}
    \hline
    Parameter & Symbol & Value\\
    \hline \hline
    Resonator frequency & $\omega_\mathrm{r}/2\pi$  & 6.538 GHz \\
    $g-e$ transition frequency & $\omega_\mathrm{ge}/2\pi$   & 4.663 GHz \\
    Anharmonicity of qubit & $\alpha/2\pi$  & -261.8 MHz \\
    Resonator--qubit detuning & $\delta_\mathrm{d}/2\pi$  & 1.876 GHz \\
    Qubit lifetime & $T_1$ & 9.6 \si{\micro\second} \\
    Qubit coherence time & $T_2^*$ & 2.3 \si{\micro\second} \\
    Josephson energy/Qubit charging energy & $E_\mathrm{J}/E_\mathrm{C}$  & 44.2 {} \\
    Qubit island capacitance & $C_\mathrm{q}$ & 73.9 fF \\
    Bare resonator relaxation rate & $\kappa_\mathrm{r}$  & \num{3.8e6} 1/s \\
    SINIS$-$resonator coupling capacitance * & $C_\mathrm{c}$ & 23.4 fF \\
    NIS junction capacitance * & $C_\mathrm{j}$ & 1.99 fF \\
    SINIS normal-metal-island capacitance * & $C_\mathrm{N}$ & 2.35 fF \\
    Tunnel resistance of SINIS & $R_\mathrm{T}$  & 44 ${\rm k\Omega}$ \\
    Dynes parameter & $\gamma_\mathrm{D}$   &\num{1.8e-4} \\
    Energy gap parameter of Al leads & $\mathit{\Delta}$ & \SI{193}{\micro\electronvolt}   \\
    
    \hline
  
\end{tabularx}

  \caption{Sample parameters used in the experiment. Each parameter is defined by a separate measurement. The parameters marked with an asterisk are design values.
  }
  
\end{table}

\newpage

\section{AC stark shift of the Qubit under the cavity drive}
\sn{To probe the QCR-induced absorption and emission of photons inside the resonator, we used a transmon qubit. The frequency shift of the qubit is the sum of Lamb shift and AC Stark shift which is proportional to the intra-resonator photon number $\langle n \rangle = \hat{a}^\dagger\hat{a}$. Figure 2 (a), (b), and (c) show the results of the qubit spectroscopy with the conventional dispersive readout after applying a 3 {\textmu}s resonator-drive pulse at various amplitudes of the resonator-drive microwave. Because the resonator’s internal energy (photon number) is proportional to the square of the amplitude of the resonator-drive microwave, the frequency of the qubit should be quadratically changed with the drive amplitude, as shown in Fig. 2 (b). The width of the resonant frequency also linearly depends on the intraresonator photon number, so it also quadratically increases the drive amplitude (Fig. 2 (c)).}

\sn{The experimental data clearly showed that the qubit resonance depends on the photon number inside the resonator, as expected from the AC stark shift. Here, the width of the resonance peak of the qubit even at the lowest resonator-drive (0.1 V) is relatively wider compared to the value estimated from the qubit coherence time $T_{1} \sim$ 9 {\textmu}s. This is because we use a relatively strong pulse for the qubit spectroscopy to ensure that the $\pi$-pulse of the qubit becomes $\sim$ 100 ns, which is shorter than the relaxation time of the resonator $\sim$ 260 ns estimated from the relaxation rate at the off state of the QCR. \sn{Due to this relaxation, the measured photon number is about 68 \% of the one immediately after the cooling completion.} Consequently, this approach enables us to detect the average of the photon number inside the resonator before the induced photons are completely depleted.}

\begin{figure}[h]
 \begin{center}
  \includegraphics[width=18cm]{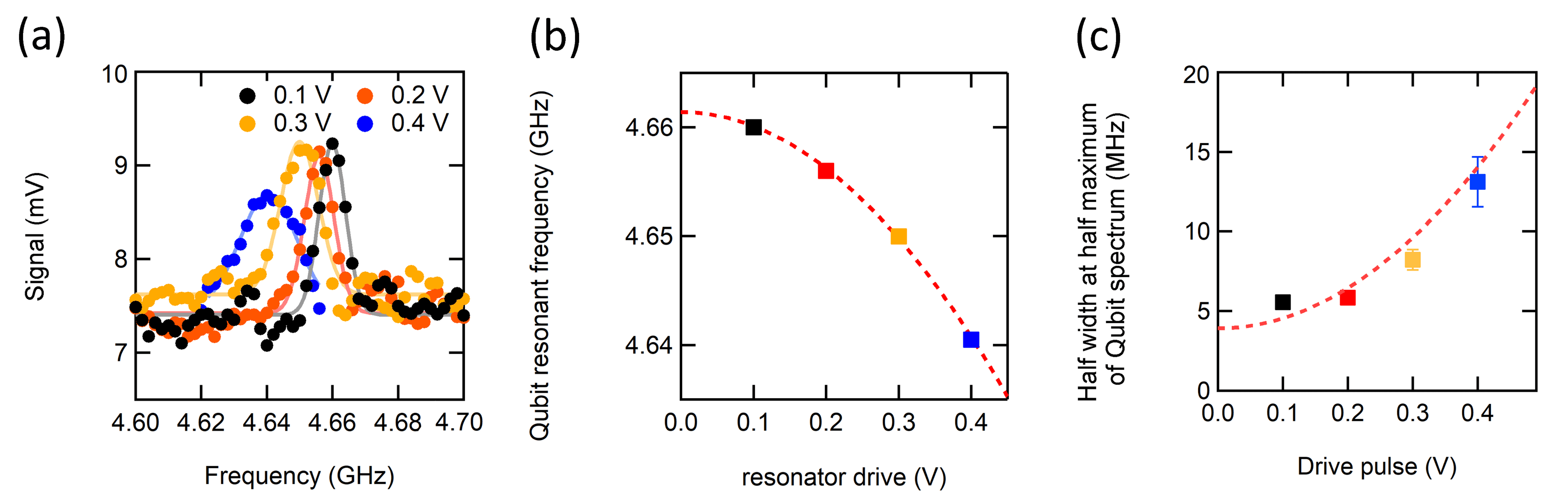}
  \caption{(a) The AC Stark shift of the qubit against various microwave pulse amplitudes, which induces the photon inside the resonator (Black: 0.1 V, Red: 0.2 V, Yellow: 0.3 V, and Blue: 0.4 V). The solid lines show the result of numerical fitting with the Gaussian function. (b)(c) The qubit resonant frequency (b) and the linewidth (c) against the microwave pulse amplitude. The red dotted line is a result of numerical fitting using the quadratic function.}
  \label{}
 \end{center}
\end{figure}

\newpage

\section{The normalized rate of quasiparticle tunneling}
\sn{In the main text, we calculated the amount of photon absorption and emission by the quantum circuit refrigerator during the square short pulse (100 ns) induced to the SINIS using the master equation. In the calculation, we utilized the normalized rate $\mathcal{F}$ of forward ($\rightarrow$) and backward ($\leftarrow$) quasiparticle tunneling in SIN junction for $R_\mathrm{T} = R_\mathrm{K}$ at the energy bias $E$}

\begin{align} \label{photon number}
    \overrightarrow{\mathcal{F}}(E) = \frac{1}{h}\int d\epsilon n_{\mathrm{S}}(\epsilon) \left[ 1 - f_{\mathrm{S}}(\epsilon)\right]f_{\mathrm{N}}(\epsilon -E)\,, 
\end{align}

\begin{align} \label{photon number2}
    \overleftarrow{\mathcal{F}}(E) = \frac{1}{h}\int d\epsilon n_{\mathrm{S}}(\epsilon) f_{\mathrm{S}}(\epsilon) \left[ 1 - f_{\mathrm{N}}(\epsilon-E)\right]\,, 
\end{align}

 \sn{When we assume that the temperature of the normal metal and superconducting lead are the same,  $\overleftarrow{\mathcal{F}}(E) = \overrightarrow{\mathcal{F}}(-E) = e^{-E/k_BT_N}\overrightarrow{\mathcal{F}}(E)$. 
In the main text, we simply describe this normalized rate $\overrightarrow{\mathcal{F}}$ as $\mathcal{F}$.}

\newpage
\section{influence of electron cooling and heating of the SIN junction \\ on the photon absorption and emission}

\sn{ In the manuscript, the experimental results deviate from the theoretical calculation, especially in the photon emission. In the theoretical calculations, we don’t take into account the following conditions to make the calculation simpler.}
\sn{
\begin{enumerate}
\item The electron heating and electron cooling which are usually observed in SIN junctions
\item Temperature differences between the normal metal and the superconducting leads in the SINIS
\item The frequency shift of the resonator due to the broadband Lamb shift \cite{Silveri2019}
\item Imperfections in the square pulse shape for SINIS
\end{enumerate}
}

\sn{Firstly, our numerical calculations do not take into account the fact that the temperature of the normal metal in SINIS junctions differs from the bath temperature due to electron cooling or heating. In the SIN junction, it is well known that when the bias voltage approaches the value of the gap, the electron temperature $T_\mathrm{N}$ becomes lower than the bath temperature $T_\mathrm{Bath}$. Conversely, when the bias voltage exceeds the gap, the electron temperature $T_\mathrm{N}$ becomes higher than the bath temperature $T_\mathrm{Bath}$. In our experiment, electron heating and cooling should play a role in the photon emission and absorption in the QCR. While it is currently challenging to instantaneously measure the temperature of the normal metal in SINIS during pulse experiments with existing technology, this remains an issue to be addressed in future research. Also, this cooling and heating lead to the temperature difference between the normal metal and superconducting leads of SINIS. This was also not taken into account. Next, our calculations do not consider the effect of changes in the resonance frequency of the superconducting resonator due to the broadband Lamb shift \cite{Silveri2019}. When we change the bias voltage to the SINIS, not only the relaxation rate but also the frequency of the resonator is changed, this frequency shift should reduce the stored photon number inside the cavity, so the experimental result should be smaller than the theoretical calculation. Finally, the actual pulse shape applied to the SINIS may deviate from a square wave due to the bandwidth limitations of amplifiers and filters, and such effects are not included in our model. To fully understand the reasons for the reduced photon emission, further theoretical and experimental investigations are necessary.}

\sn{Here, we focus on the effect of electron cooling and heating in the SIN junctions on the total number of photons inside the cavity more in detail. To investigate the effect of this electron cooling and heating on the photon absorption and emission, we numerically calculated the photon absorption and emission under the condition that the electron temperature of normal metal $T_{N}$ deviates from the temperature of the bath $T_{Bath}$.}

\begin{figure}[h]
 \begin{center}
  \includegraphics[width=15cm]{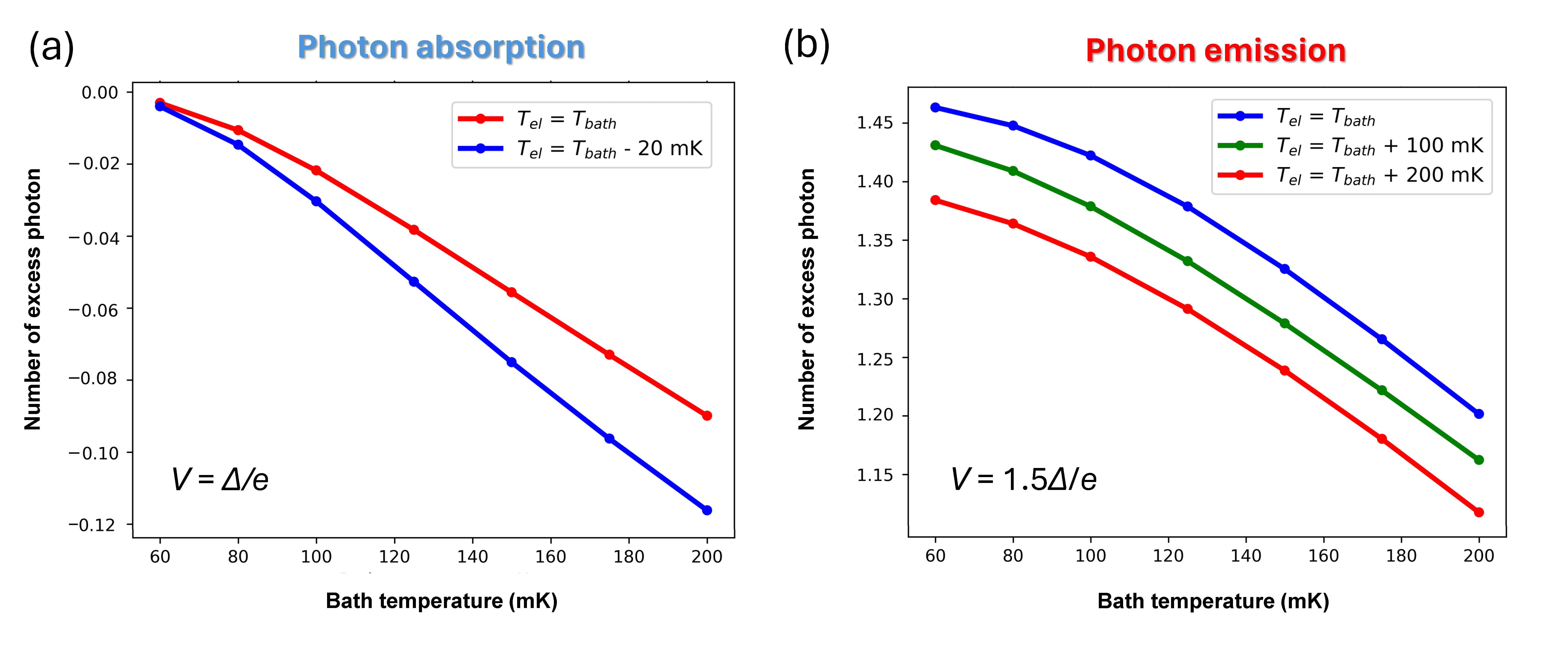}
  \caption{The number of excess photons inside the resonator when the electron temperature of the normal metal $T_\mathrm{N}$ deviates from the bath temperature $T_\mathrm{Bath}$. (a) Number of excess photons in the condition $V_{SINIS} =\Delta/e$. The red curve shows the result with $T_\mathrm{N} = T_\mathrm{Bath}$ (no cooling) and the blue curve shows the results with $T_\mathrm{N} = T_\mathrm{Bath} -20 mK$ (electron cooling). (b) Number of excess photons in the condition $V_\mathrm{SINIS} =1.5 \Delta/e$. The blue curve shows the result with $T_\mathrm{N} = T_\mathrm{Bath}$ (no heating) and the green and the red curve shows the results with $T_\mathrm{N} = T_\mathrm{Bath}$ + 100 mK and $T_\mathrm{Bath}$ + 200 mK (electron heating)  }
  \label{}
 \end{center}
\end{figure}

\sn{Figure 3 (a) shows the result of the numerical calculation for the number of excess photons when $V_\mathrm{SINIS} = \Delta/e$. As shown in the main text, it clearly showed that the amount of photon absorption becomes larger as the "bath" temperature increases. Furthermore, assuming that the “electron" temperature of the normal metal is colder than the bath temperature ($T_\mathrm{N} = T_\mathrm{Bath}$ - 20 mK) due to the electron cooling, numerical calculations indicate an increase in photon absorption.}

\sn{On the other hand, in regions where the bias voltage exceeds the superconducting gap, photon emission becomes dominant and the temperature of the normal metal in SINIS structures exceeds the bath temperature. When we assume that the temperature of the normal metal is higher while keeping the bath temperature constant, calculations reveal that the photon emission is reduced (approximately 10\%) compared to the one where the temperature of the normal metal is set to be the bath temperature (Fig. 3 (b)).}

\sn{The actual electron temperature of the normal metal depends on device structure, and the actual temperature of the normal metal is impossible to be measured instantaneously with the current temperature measurement technique. From the above numerical calculations, we can consider that the discrepancy between the theoretical calculations and the experiments regarding photon emission likely arises from a complex interplay of several factors mentioned above. }


\newpage
\section{The absolute number of photons inside the cavity before and after quantum circuit refrigeration}

\sn{Here, we discuss the absolute number of photons before and after cooling from equilibrium states at various bath temperatures ($T_\mathrm{Bath}$ = 15 mK, 100 mK, 150 mK, and 200 mK). }

\sn{First, in this study, because the temperature of the substrate (with superconducting resonator, superconducting qubit, and SINIS) was not thermalized to the bath temperature ($T_\mathrm{Bath}$ = 15 mK), it was necessary to measure the substrate temperature using an alternative method. Here, we estimated the temperature using Rabi population measurement\cite{Jin 2015} with a superconducting qubit. In thermal equilibrium, the occupancy of the excited state of the qubit is determined by the Boltzmann distribution, so from the occupancy of the excited state, we can estimate the temperature. This thermometry should directly measure the electron temperature of the substrate which has a qubit, providing a more accurate representation of the device temperature than the Ruthenium oxide thermometer on the mixing chamber plate in the dilution refrigerator. The results of the Rabi population measurement at the lowest temperature of the dilution refrigerator ($T_\mathrm{Bath}$ = 15 mK) are shown in Fig. 4 . From the Rabi population measurement, we obtain the excited state population of $\sim$ 16$\%$, which corresponds to a temperature $T_{device}$ = 120 mK. This value should be considered to be closer to the actual device temperature. From this device temperature, we determined that the average photon occupancy in the resonator at the lowest bath temperature is 0.07. }

\begin{figure}[h]
 \begin{center}
  \includegraphics[width=15cm]{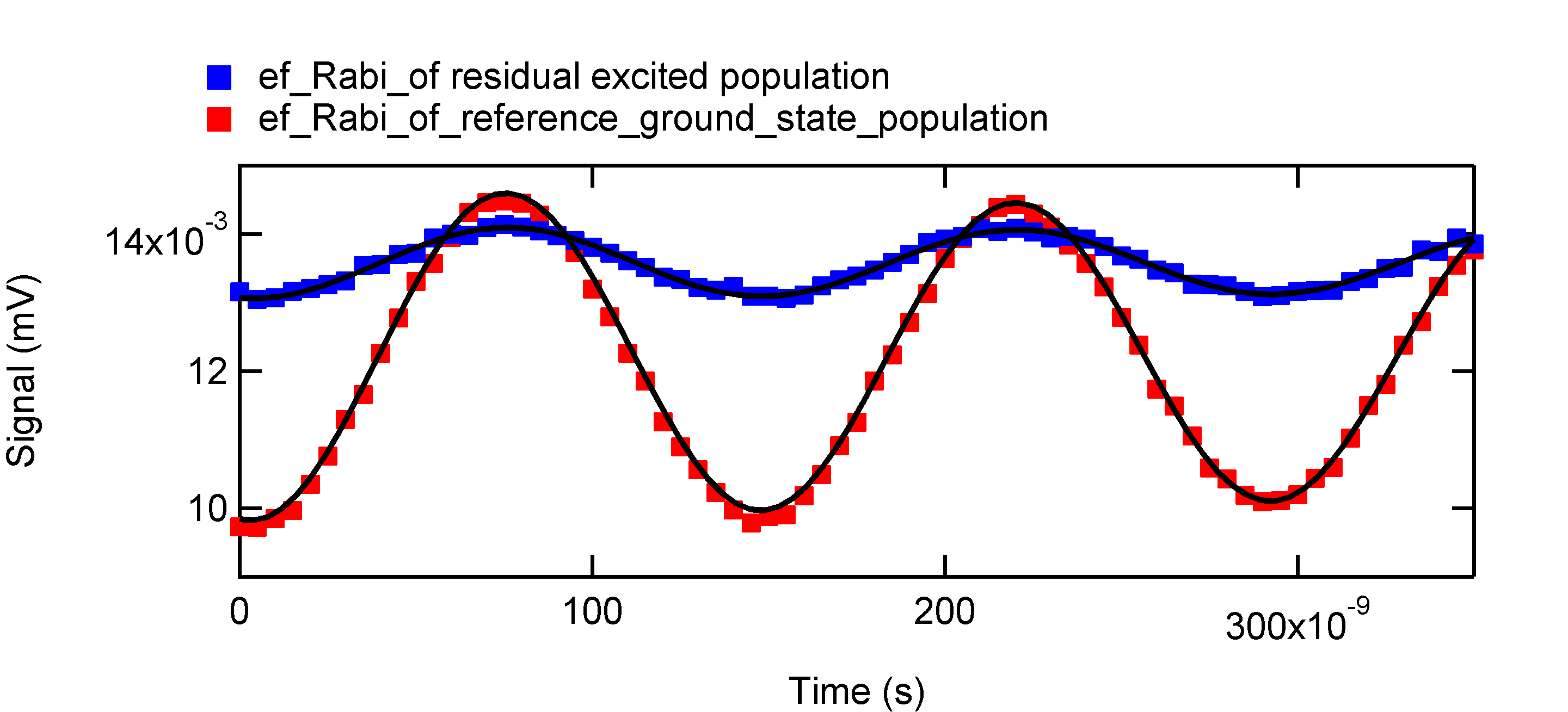}
  \caption{Rabi population measurement at the lowest bath temperature of the dilution refrigerator ($T_\mathrm{Bath} =$ 15 mK)}
  \label{}
 \end{center}
\end{figure}

\sn{Next, from the measurements conducted for Fig. 3 in the main text, we can determine the extent to which the resonance frequency of the qubit at each temperature ($T_\mathrm{Bath}$ = 100 mK, 150 mK, 200 mK) (under no cavity drive and no SINIS bias voltage $V_\mathrm{SINIS}$) is shifted compared to the resonance frequency at the lowest bath temperature in the same condition (no cavity drive and no SINIS bias voltage$V_\mathrm{SINIS}$). This shift is caused by the AC Stark shift due to thermally excited photons. From this, we can estimate the increase in photon number with temperature. Figure 5 shows the frequency shift compared to the one at the base temperature of the dilution refrigerator. The number of thermally excited photons in the resonator increases with increasing temperature.}

\sn{Finally, from Fig. 3 (c) in the main text, we know the amount of photon absorption at each temperature from the AC Stark shift. Consequently, we can estimate the number of photons in the equilibrium state before and after cooling at each temperature. As is evident from the graph, our quantum circuit refrigeration operates in the regime below the single energy quantum level (Photon number $\ll$ 1).}

\begin{figure}[h]
 \begin{center}
  \includegraphics[width=15cm]{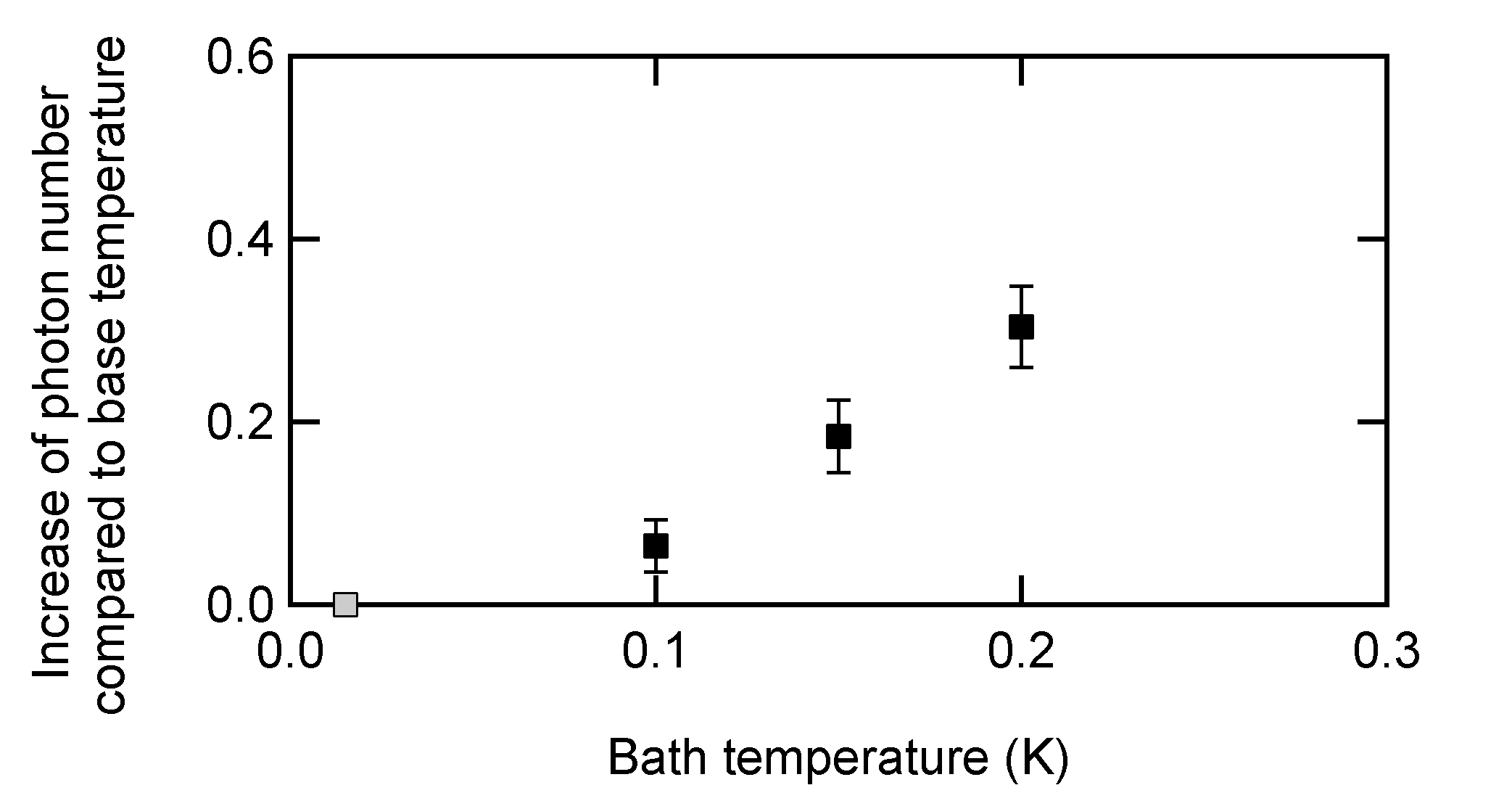}
  \caption{Increase of photon number compared to that at the base temperature ($T_\mathrm{Bath}$ = 15 mK,)}
  \label{}
 \end{center}
\end{figure}

\begin{figure}[h]
 \begin{center}
  \includegraphics[width=15cm]{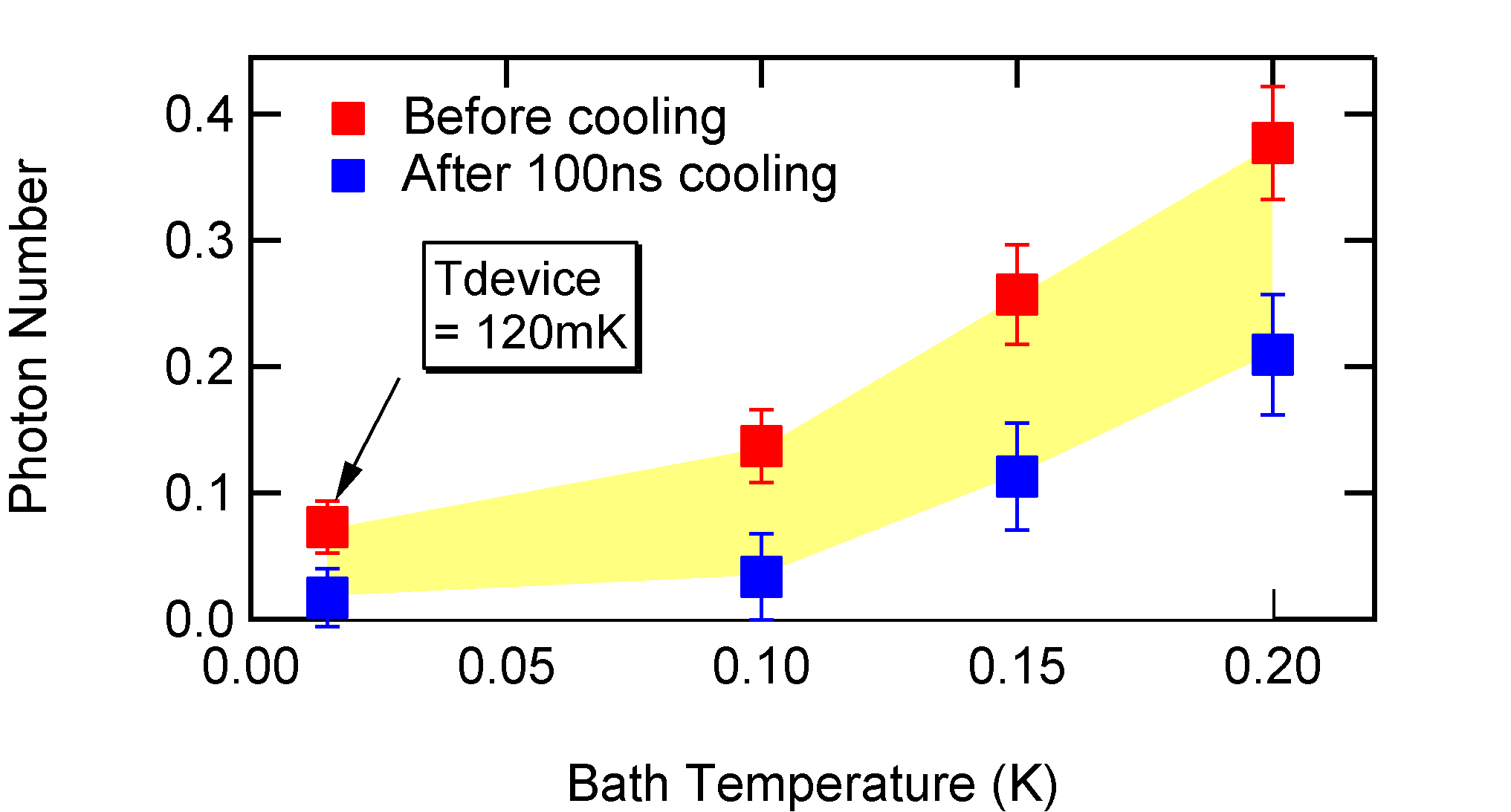}
  \caption{Absolute value of photon number inside the cavity before (red squares) and after (blue squares) 100 ns cooling pulse at various bath temperatures ($T_\mathrm{Bath}$ = 15 mK, 100mK, 150 mK, and 200 mK)}
  \label{}
 \end{center}
\end{figure}